\documentstyle[aps,psfig,prd]{revtex}

\begin{document}
\hbadness=10000 \pagenumbering{arabic}

\preprint{{\vbox{\hbox{
IPAS-HEP-01-k003, NCKU-HEP-01-06}}}} \vspace{1.5cm}

\title{\Large \bf
Three-body nonleptonic $B$ decays in perturbative QCD
}
\date{\today}
\author{\large \bf  Chuan-Hung Chen\footnote{Email:
chchen@phys.sinica.edu.tw} and Hsiang-nan
Li\footnote{Email: hnli@phys.sinica.edu.tw}}

\vskip1.0cm

\address{ Institute of Physics, Academia Sinica,
Taipei, Taiwan 115, Republic of China}
\maketitle

\begin{abstract}

We develop perturbative QCD formalism for three-body nonleptonic $B$
meson decays. Leading contributions are identified by defining power
counting rules for various topologies of amplitudes. The analysis is
simplified into the one for two-body decays by introducing two-meson
distribution amplitudes. This formalism predicts both nonresonant and
resonant contributions, and can be generalized to baryonic decays.

\end{abstract}

\vskip 1.0cm

The fundamental concept of perturbative QCD (PQCD) is to separate hard
and soft dynamics in a QCD process. The former is calculable in
perturbation theory, while the latter, though not calculable, is
treated as a universal input. The separation can be performed in the
framework of collinear factorization \cite{BL} or of $k_T$ factorization
\cite{BS,NL}, in which
an amplitude is expressed as a convolution of a hard kernel $H$ with a
hardon distribution amplitude $\Phi(x)$ or with a hadron wave function
$\Phi(x,k_T)$, $x$ and $k_T$ being a longitudinal momentum fraction and
a transverse momentum, respectively. Collinear factorization works,
if it does not develop an end-point singularity from $x\to 0$ in the
above convolution. If it does, collinear factorization breaks down, and
$k_T$ factorization is more appropriate.

It has been known that collinear factorization of charmed and charmless
two-body $B$ meson decays suffers end-point singularities. The PQCD
formalism for these modes based on $k_T$ factorization theorem was then
derived \cite{YL,KLS,LUY}, which has been shown to be
infrared-finite, gauge-invariant, and consistent with the factorization
assumption in the heavy-quark limit \cite{TLS,LU}. If one still employs
collinear factorization, an alternative approach, the so-called
QCD-improved factorization \cite{BBNS}, can be developed. In this
approach the end-point singularities in the leading contributions are
absorbed into $B$ meson transition form factors, and those appearing at
the subleading level are regularized by arbitrary (nonuniversal)
infrared cutoffs of momentum fractions $x$. Without the arbitrary
cutoffs, PQCD has a predictive power, whose predictions for $B\to PP$,
$VP$, and $VV$ modes are all in agreement with data \cite{Keum}.


Three-body nonleptonic $B$ meson decays have been observed recently
\cite{Belle,Bar}. Viewing the experimental progress, it is urgent to
construct a corresponding framework. Motivated by its theoretical
self-consistency and phenomenological success, we shall generalize PQCD
to these modes. A direct evaluation of the hard kernels, which contain
two virtual gluons at lowest order, is not practical due to the enormous
number of diagrams. On the other hand, the region with the two gluons
being hard simultaneously is power-suppressed and not important.
Therefore, a new input is necessary in order to catch dominant
contributions to three-body decays in a simple manner. The idea is to
introduce two-meson distribution amplitudes \cite{MP}, by means of which
a factorization formula for a $B\to h_1h_2h_3$ decay amplitude is
written, in general, as
\begin{eqnarray}
{\cal M}=\Phi_B\otimes H\otimes \Phi_{h_1h_2}\otimes\Phi_{h_3}\;.
\end{eqnarray}
It will be shown that both nonresonant contributions
and resonant contributions through two-body channels can be included
through the parametrization of the two-meson distribution amplitude
$\Phi_{h_1h_2}$.

Three-body decay amplitudes are classified into four topologies,
depending on number of light mesons emitted from the four-fermion
vertices. Topologies I and III, shown in Figs.~1(a) and 1(c), are
associated with one light meson emission and three light meson emission,
respectively. The bubbles denote the distribution amplitudes, which
absorb nonperturbative dynamics. The hard kernel $H$ contains only a
single hard gluon exchange. The former involves transition of the $B$
meson into two light mesons. In the latter case a $B$ meson annihilates
completely. For two light meson emission shown in Fig.~1(b), we assign IIs
to the special amplitude corresponding to the scalar vertex, and II to the
rest of the amplitudes. Both topologies II and IIs are expressed as a
product of a heavy-to-light form factor and a time-like light-light form
factor in the heavy-quark limit.

The dominant kinematic region for three-body $B$ meson decays is the one,
where at least one pair of light mesons has the invariant mass of
$O(\bar\Lambda M_B)$ for nonresonant contributions and of
$O(\bar\Lambda^2)$ for resonant contributions, $\bar\Lambda=M_B-m_b$
being the $B$ meson and $b$ quark mass difference. An example is
the configuration, where all three mesons carry momenta of $O(M_B)$, but
two of them move almost parallelly. In the above dominant region
collinear factorization theorem applies to topology I, since it is free
of end-point singularities as shown below. With the pair of mesons emitted
with a small invariant mass, the evaluation of topologies II and IIs is
the same as of two-body decays. The contribution from the region,
where all three pairs have the invariant mass of $O(M_B^2)$, is
power-suppressed. This contribution is the one, which can be calculated
perturbatively in terms of the diagrams with two hard gluon exchanges.

We define the power counting rules for the various topologies in the
dominant kinematic region, and identify the leading ones. Consider first
nonresonant contributions. Topology I behaves like
$(\bar\Lambda M_B)^{-2}$, where one power of $(\bar\Lambda M_B)^{-1}$
comes from the hard gluon in Fig.~1(a), which kicks the soft spectator
in the $B$ meson into a fast one in a light meson \cite{TLS}, and another
power is attributed to the invariant mass of the light meson pair. The
overall product of the meson decay constants
is not shown explicitly. Topology II exhibits the same power behavior as
topology I: the hard gluon in Fig.~1(b), i.e., the $B$ meson transition
form factor, gives a power of $(\bar\Lambda M_B)^{-1}$, and the
light-light form factor gives another power. The scalar vertex introduces
an extra power $m_0/M_B$, $m_0$ being the chiral symmetry breaking scale,
to topology IIs. Topology III must involve large energy release for
producing at least a pair of fast mesons with the invariant mass of
$O(M_B^2)$. That is, it behaves like $(\bar\Lambda M_B)^{-1}M_B^{-2}$.
Hence, we have the relative importance of the decay amplitudes,
\begin{eqnarray}
{\cal M}_{\rm I}:{\cal M}_{\rm II}:{\cal M}_{\rm IIs}:{\cal M}_{\rm III}
=1:1:\frac{m_0}{M_B}:\frac{\bar\Lambda}{M_B}\;,
\label{pow}
\end{eqnarray}
indicating that topology III is negligible. For resonant contributions,
we replace the power of $(\bar\Lambda M_B)^{-1}$ associated with
the light meson pair by $\bar\Lambda^{-2}$. Therefore,
Eq.~(\ref{pow}) still holds.

Take topology I for the $B^+\to K^+\pi^+\pi^-$ mode as an example, in
which the $B$ meson transit into a pair of pions. The $\pi^+$ and
$\pi^-$ mesons carry the momenta $P_1$ and $P_2$, respectively.
The $B$ meson momentum $P_B$, the total momentum of the two pions,
$P=P_1+P_2$, and the kaon momentum $P_3$ are chosen as
\begin{eqnarray}
P_B=\frac{M_B}{\sqrt{2}}(1,1,{\bf 0}_T)\;,\;\;\;\;
P=\frac{M_B}{\sqrt{2}}(1,\eta,{\bf 0}_T)\;,\;\;\;\;
P_3=\frac{M_B}{\sqrt{2}}(0,1-\eta,{\bf 0}_T)\;,
\end{eqnarray}
with the variable $\eta=w^2/M_B^2$, $w^2=P^2$ being the invariant mass
of the two-pion system. The light-cone coordinates have been adopted here.
Define $\zeta=P_1^+/P^+$ as the $\pi^+$ meson momentum fraction, in
terms of which, the other kinematic variables are expressed as
\begin{eqnarray}
P_2^+=(1-\zeta)P^+\;,\;\;\;\;
P_1^-=(1-\zeta)\eta P^+\;,\;\;\;\;
P_2^-=\zeta\eta P^+\;,\;\;\;\;
P_{1T}^2=P_{2T}^2=\zeta(1-\zeta)w^2\;.
\end{eqnarray}

The two pions from the $B$ meson transition possess the invariant mass
$w^2\sim O(\bar\Lambda M_B)$, implying the orders of magnitude
$P^+\sim O(M_B)$, $P^-\sim O(\bar\Lambda)$ and
$P_T\sim O(\sqrt{\bar\Lambda M_B})$.
In the heavy-quark limit, the hierachy $P^+\gg P_{1(2)T}\gg P^-$
corresponds to a collinear configuration.
Therefore, we introduce the two-pion distribution amplitudes \cite{MP},
\begin{eqnarray}
\Phi_v(z,\zeta,w^2)&=&\frac{1}{2\sqrt{2N_c}}\int \frac{dy^-}{2\pi}
e^{-izP^+y^-}\langle\pi^+(P_1)\pi^-(P_2)|\bar\psi(y^-)\not n_-T
\psi(0)|0\rangle\;,
\label{pa}\\
\Phi_s(z,\zeta,w^2)&=&\frac{1}{2\sqrt{2N_c}}\frac{P^+}{w}
\int \frac{dy^-}{2\pi}
e^{-izP^+y^-}\langle\pi^+(P_1)\pi^-(P_2)|\bar\psi(y^-)T
\psi(0)|0\rangle\;,
\label{ps}\\
\Phi_t(z,\zeta,w^2)&=&\frac{1}{2\sqrt{2N_c}}\frac{f_{2\pi}^\perp}{w^2}
\int \frac{dy^-}{2\pi}
e^{-izP^+y^-}\langle\pi^+(P_1)\pi^-(P_2)|\bar\psi(y^-)
i\sigma_{\mu\nu}n_-^\mu P^\nu T\psi(0)|0\rangle\;,
\label{pt}
\end{eqnarray}
with $\Phi_v$ being the twist-2 component, and $\Phi_s$ and $\Phi_t$
the twist-3 components. $T=\tau^3/2$ is for the isovector $I=1$
state, $\psi$ the $u$-$d$ doublet, $z$ the momentum fraction carried by
the spectator $u$ quark, and $n_-=(0,1,{\bf 0}_T)$ a dimensionless vector.
The constant $f_{2\pi}^\perp$ of dimension of mass is defined via
the local matrix element \cite{MP},
\begin{eqnarray}
\lim_{w^2\to 0}\langle\pi^+(P_1)\pi^-(P_2)|\bar\psi(0)
i\sigma_{\mu\nu}n_-^\mu P^\nu \frac{\tau^3}{2}\psi(0)|0\rangle
=\frac{w^2}{f_{2\pi}^\perp}(2\zeta-1)P^+\;.
\end{eqnarray}

The matrix element with the
structure $\gamma_5\not n_-$ vanishes for topologies I and IIs, and
contributes to topology II at twist 4. The one with the structure
$\gamma_5$ vanishes. For topologies II and IIs, a kaon-pion distribution
amplitude is introduced in a similar way.
For other two-pion systems, the distribution amplitudes can be defined
with the appropriate choice of the matrix $T$. For instance, $T=1/2$ is
for the $\pi^0\pi^0$ isoscalar ($I=0$) state.

A two-pion distribution amplitude can be related to the pion distribution
amplitude through the calculation of the process
$\gamma\gamma^*\to \pi^+\pi^-$ at large invariant mass $w^2$ \cite{DFK}.
The extraction of the two-pion distribution amplitudes from the
$B\to \pi\pi l\bar\nu$ decay has been discussed in \cite{M}. Here
we pick up the leading term in the complete Gegenbauer expansion of
$\Phi_i(z,\zeta,w^2)$ \cite{MP}:
\begin{eqnarray}
\Phi_{v,t}(z,\zeta,w^2)=\frac{3F_{\pi,t}(w^2)}{\sqrt{2N_c}}
z(1-z)(2\zeta-1)\;,
\;\;\;\;
\Phi_{s}(z,\zeta,w^2)=\frac{3F_{s}(w^2)}{\sqrt{2N_c}}z(1-z)\;,
\label{2pi}
\end{eqnarray}
where $F_{\pi,s,t}(w^2)$ are the time-like pion electromagnetic, scalar
and tensor form factors with $F_{\pi,s,t}(0)=1$. That is, the two-pion
distribution amplitudes are normalized to the time-like form factors.
For $\Phi_t$ in Eq.~(\ref{2pi}), we have adopted the parametrization,
\begin{eqnarray}
\langle\pi^+(P_1)\pi^-(P_2)|\bar\psi(0)
i\sigma_{\mu\nu}n_-^\mu P^\nu \frac{\tau^3}{2}\psi(0)|0\rangle
=\frac{w^2}{f_{2\pi}^\perp}F_t(w^2)(2\zeta-1)P^+\;.
\end{eqnarray}
Note that the asymptotic functional form for the $z$ dependence of
$\Phi_s$ is an assumption.

For the $B$ meson distribution amplitude, we employ the model \cite{KLS},
\begin{eqnarray}
\Phi_{B}(x)=N_Bx^2(1-x)^2\exp
\left[-\frac{1}{2}\left(\frac{xM_B}{\omega_B}\right)^2\right]\;,
\label{phib}
\end{eqnarray}
with the shape parameter $\omega_{B}=0.4$ GeV, and the normalization
constant $N_{B}$ related to the decay constant $f_{B}=190$ MeV (in the
convention $f_{\pi}=130$ MeV) via
$\int_{0}^{1}\Phi_{B}(x)dx=f_{B}/(2\sqrt{2N_c})$. The above $\Phi_B$ is
identified as $\Phi_+$ in the definition of the two leading-twist $B$
meson distribution amplitudes $\Phi_\pm$ given in \cite{GN,DS}.
Equation (\ref{phib}), vanishing at $x\to 0$, is consistent with the
behavior required by equations of motion \cite{KKQT}. Another
distribution amplitude ${\bar\Phi}_B$ in our definition, identified
as ${\bar\Phi}_B=(\Phi_B^{-}-\Phi_B^{+})/\sqrt{2}$
with a zero normalization, contributes at the next-to-leading power
$\bar\Lambda/M_B$ \cite{TLS}. It has been verified numerically \cite{LMY}
that the contribution to the $B\to\pi$ form factor from $\Phi_B$ is much
larger than from ${\bar\Phi}_B$.

The total decay rate is written as
\begin{eqnarray}
\Gamma=\frac{G_F^2M_B^5}{512\pi^4}\int_0^1 d\eta (1-\eta)\int_0^1 d\zeta
|{\cal M}|^2\;,\;\;\;\;{\cal M}={\cal M}_{\rm I}+{\cal M}_{\rm II}+
{\cal M}_{\rm IIs}\;,
\end{eqnarray}
with the amplitudes,
\begin{eqnarray}
{\cal M}_{\rm I}&=&f_{K}\Big(V^{*}_{t}\sum_{i=4,6}{\cal F}^{P(u)}_{ei}
- V^{*}_{u}{\cal F}_{e2}\Big)\;, \;\;\;\;
{\cal M}_{\rm IIs}=V^{*}_{t}F_{s}(\omega^2)F^{P(d)}_{e6} \;,
\nonumber \\
{\cal M}_{\rm II}&=&
(2\zeta-1)F_{\pi}(\omega^2)\Big[V^{*}_{t}\Big(\sum_{i=3}^5 F^{P(d)}_{ei}
+\sum_{i=3,5}F^{P(u)}_{ei}\Big)-V^{*}_{u} F_{e1}\Big]\;.
\end{eqnarray}
For a simpler presentation, we have assumed that the kaon-pion time-like
form factor in topology IIs is equal to the pion time-like form factor
multiplied by the ratio of the decay constants $f_K/f_\pi$. This
assumption is in fact not necessary, and the property of the kaon-pion
form factor will be discussed elsewhere. The superscript $P(q)$ stands
for the amplitude from a penguin operator producing a pair of quarks $q$.
Those without $P(q)$ arise from tree operators. The subscript $ei$ stands
for the emission topology (in contrast to the annihilation topology III)
from the effective four-fermion operator $O_i$ in the standard notation.

We calculate the hard kernels by contracting the structures, which
follow Eqs.~(\ref{pa})-(\ref{pt}),
\begin{equation}
\frac{(\not{P}_B+M_B)\gamma_5}{\sqrt{2N_c}}\Phi_B(x)\;,\;\;\;\;
\frac{1}{\sqrt{2N_c}}\left[\not P\Phi_v(z,\zeta,w^2)
+w\Phi_s(z,\zeta,w^2)
+\frac{\not P_1\not P_2-\not P_2\not P_1}{w(2\zeta-1)}
\Phi_t(z,\zeta,w^2)\right]\;,
\end{equation}
to Fig.~1. The factorization formulas for the $B\to 2\pi$ transition
amplitudes are given by
\begin{eqnarray}
{\cal F}^{P(u)}_{e4}&=& 8\pi C_{F} M^2_{B} (1-\eta) \int^{1}_{0}
dx_{1} dz\frac{\Phi_{B}(x_{1})}{x_{1}zM^{2}_{B}+P_{T}^2}
\nonumber \\
&& \times \Big\{\Big[
(1+z)\Phi_{v}(z,\zeta,w^2)+\sqrt{\eta}(1-2z)\Phi_{t}(z,\zeta,w^2)
+\sqrt{\eta}(1-2z)\Phi_{s}(z,\zeta,w^2)\Big]
\frac{\alpha _{s}( t^{(1)}_{e})a_{4}^{(u)}(t^{(1)}_{e})}
{zM^2_{B}+P_{T}^2}
\nonumber \\
&&-\Big[ \eta \Phi_{v}(z,\zeta,w^2)
-2\sqrt{\eta}\Phi_{s}(z,\zeta,w^2)\Big]
\frac{\alpha _{s}( t^{(2)}_{e})a_{4}^{(u)}(t^{(2)}_{e})}
{x_{1}M^2_{B}}\Big\}\;,
\nonumber\\
{\cal F}^{P(u)}_{e6}&=& -16\pi C_{F} M^2_{B} r_{0} \int^{1}_{0}
dx_{1} dz\frac{\Phi_{B}(x_{1})}{x_{1}zM^{2}_{B}+P_{T}^2}
\nonumber \\
&& \times \Big\{\Big[ (1+\eta-2z\eta)\Phi_{v}(z,\zeta,w^2)
-\sqrt{\eta} z\Phi_{t}(z,\zeta,w^2)
+\sqrt{\eta}(2+z)\Phi_{s}(z,\zeta,w^2)\Big]
\frac{\alpha _{s}( t^{(1)}_{e})a_{6}^{(u)}(t^{(1)}_{e})}
{zM^2_{B}+P_{T}^2}
\nonumber\\
&&-\Big[ \eta\Phi_{v}(z,\zeta,w^2)-2\sqrt{\eta}
 \Phi_{s}(z,\zeta,w^2) \Big]
\frac{\alpha _{s}( t^{(2)}_{e})a_{6}^{(u)}(t^{(2)}_{e})}
{x_{1}M^2_{B}}\Big\}\;,
\label{cf}
\end{eqnarray}
with $r_0=m_0/M_B$. ${\cal F}_{e2}$ is the same as ${\cal F}^{P(u)}_{e4}$
but with $a^{(u)}_{4}$ replaced by $a_{2}$ (here $a_2$ is close to
unity). The definitions of the Wilson coefficients $a^{(q)}(t)$ are
referred to \cite{CKL1}. The hard scales are defined by
$t_e^{(1)}=\max[\sqrt{z}M_B, P_T]$ and
$t_e^{(2)}=\max[\sqrt{x_1}M_B, P_T]$.
The above collinear factorization formulas are well-defined,
since the invariant mass of the two-pion system,
proportional to $P_T$, smears the end-point
singularities from $z\to 0$.


The $B$ meson transition form factors involved in topologies II and IIs
are
\begin{eqnarray}
F^{P(d)}_{e4} &=&8\pi C_{F}M_{B}^{2}
\int_{0}^{1}dx_1dx_3\int_{0}^{\infty}b_{1}db_{1}b_{3}db_{3}
\Phi_{B}( x_{1},b_{1})
\nonumber \\
&&\times \Big\{ \Big[ (1-\eta)(1+(1-\eta)x_{3})\Phi_{K}(x_{3})
+r_{0}(1+\eta-2(1-\eta)x_{3})\Phi_{K}^{p}(x_{3})
\nonumber \\
&&+r_{0}(1-\eta)(1-2x_{3})\Phi_{K}^{\sigma}(x_{3})\Big]
E_4^{(d)}( t_{e}^{( 1) }) h_e(x_{1},(1-\eta)x_{3},b_{1},b_{3})
\nonumber \\
&& +2r_{0}(1-\eta)\Phi_{K}^{p}(x_{3}) E_4^{(d)}( t_{e}^{( 2) })
h_e((1-\eta)x_{3},x_{1},b_{3},b_{1}) \Big\} \;,
\nonumber\\
F^{P(d)}_{e6} &=&16\pi
C_{F}M_{B}^{2}\sqrt{\eta}\int_{0}^{1}dx_1dx_3\int_{0}^{\infty}
b_{1}db_{1}b_{3}db_{3}\ \Phi_{B}( x_{1},b_{1})
\nonumber \\
&&\times \Big\{ \Big[ (1-\eta)\Phi_{K}(x_{3})
+2r_{0}\Phi_{K}^{p}(x_{3})+r_{0}(1-\eta)x_{3}\Big(\Phi_{K}^{p}(x_{3})
-\Phi_{K}^{\sigma}(x_{3})\Big)\Big]
\nonumber \\
&&\times E_6^{(d)}( t_{e}^{( 1) }) h_e(x_{1},(1-\eta)x_{3},b_{1},b_{3})
\nonumber \\
&& +2r_{0}(1-\eta)\Phi _{K}^{p}(x_{3}) E_6^{(d)}( t_{e}^{( 2) })
h_e((1-\eta)x_{3},x_{1},b_{3},b_{1}) \Big\} \;.
\end{eqnarray}
The definitions of the evolution factors $E_i^{(q)}(t)$, which contain the
Wilson coefficients $a^{(q)}_{i}(t)$, of the hard functions $h_{e}$, and
of the kaon distribution amplitudes $\Phi_{K}$, $\Phi_{K}^{p}$ and
$\Phi_{K}^{\sigma}$ are referred to \cite{CKL1}.
$F^{P(q)}_{e3}$, $F^{P(q)}_{e5}$  and $F_{e1}$ are obtained from
$F^{P(d)}_{e4}$ by substituting $a^{(q)}_{3}$, $a^{(q)}_{5}$ and $a_1$
for $a^{(d)}_{4}$, respectively.

The PQCD evaluation of the form factors indicates the power behavior
in the asymptotic region, $F_\pi(w^2)\sim 1/w^2$, and their relative
importance: $F_{s,t}(w^2)/F_\pi(w^2)\sim m_0/w$. Therefore, the twist-3
contributions in Eq.~(\ref{cf}) are down by a power of
$\sqrt{\eta}m_0/w=m_0/M_B$
compared to the twist-2 ones, which is the accuracy considered here.
To calculate the nonresonant contribution, we propose the parametrization
for the whole ragne of $w^2$,
\begin{eqnarray}
F^{(nr)}_\pi(w^2)=\frac{m^2}{w^2+m^2}\;,\;\;\;\;
F^{(nr)}_{s,t}(w^2)=\frac{m_0 m^2}{w^3+m_0 m^2}\;,
\label{non}
\end{eqnarray}
where the the parameter $m=1$ GeV is determined by the fit to
the  experimental data $M_{J/\psi}^2|F_\pi(M_{J/\psi}^2)|^2\sim 0.9$
GeV$^2$ \cite{PDG}, $M_{J/\psi}$ being the $J/\psi$ meson mass. These
form factors can carry strong phases, which are
assumed to be not very different, {\it i.e.}, overall and negligible here.

To calculate the resonant contribution, we parametrize
it into the time-like form factors,
\begin{eqnarray}
F^{(r)}_{\pi,s,t}(w^2)=\frac{M_V^2}{\sqrt{(w^2-M_V^2)^2+\Gamma_V^2w^2}}
-\frac{M_V^2}{w^2+M_V^2}\;,
\label{res}
\end{eqnarray}
with $\Gamma_V$ being the width of
the meson $V$.  The subtraction term renders Eq.~(\ref{res}) exhibit
the features of resonant contributions: the normalization
$F^{(r)}_\pi(0)=0$ and the asymptotic behavior
$F^{(r)}_\pi(w^2)\sim 1/w^4$, which decreases at large $w$ faster than
the nonresonant parametrization in Eq.~(\ref{non}). Equation~(\ref{res})
is motivated by the pion time-like form factor measured at the $\rho$
resonance \cite{RR}. It is likely that all
$F^{(r)}_{\pi,s,t}$ contain the similar resonant contributions.
The relative phases among different resonances will be
discussed elsewhere by employing the more sophisticated
parametrization \cite{GS}. Here we assume the absence of the interference
effect.

We adopt $m_0=$ 1.4 (1.7) GeV for the pion (kaon) and the unitarity angle
$\phi_3=90^o$ \cite{KLS}. For the $B^+\to\rho^0(770) K^+$
and $B^+\to f_0(980) K^+$ channels, we choose $\Gamma_\rho=150$ MeV
and $\Gamma_{f_0}=50$ MeV \cite{KH}.
The nonresonant contribution $0.61\times 10^{-6}$ to the $B^+\to
K^+\pi^+\pi^-$ branching ratio is obtained. Our results $1.8\times
10^{-6}$ and $13.2\times 10^{-6}$ are consistent with the measured
three-body decay branching ratios through the $B^+\to \rho(770)
K^+$  and $B^+\to f_0(980) K^+$ channels, $< 12\times 10^{-6}$ and
$(9.6^{+2.5+1.5+3.4}_{-2.3-1.5-0.8})\times 10^{-6}$ \cite{Belle},
respectively. Since the $f_0$ width has a large uncertainty, we
also consider $\Gamma_{f_0}=60$ MeV, and the branching ratio
reduces to $10.5\times 10^{-6}$. The resonant contributions from
the other channels can be analyzed in a similar way. For example,
the $K^*(892)$ resonance can be included into the $K$-$\pi$ form
factors by choosing the width $\Gamma_{K^*}=50$ MeV. The
nonresonant and resonant contributions to the $B^+\to
K^+\pi^+\pi^-$ decay spectrum are displayed in Fig.~2.


In the above formalism nonfactorizable contributions arise from the
diagrams, in which a hard gluon attaches the spectator quark and the meson
emitted from the weak vertex (topology I) or the meson pair (topologies
II and IIs). The nonfactorizable contributions, suppressed by
$\ln^{-1}(M_B/\bar\Lambda)$ \cite{LU}, and topology III, being of
$O(\bar\Lambda/M_B)$, can be evaluated systematically by means of
the two-meson distribution amplitudes. The framework presented here is
not only applicable to the study of three-body mesonic $B$ meson decays,
but also to baryonic decays \cite{CHT}, such as $B\to p\bar p K$. One
simply introduces two-proton distribution amplitudes, and the calculation
of the corresponding hard kernel is similar.

In this letter we have proposed a promising formalism for three-body
nonleptonic $B$ meson decays. This formalism, though at its early stage,
is general enough for evaluating both nonresonant and resonant
contributions to various modes, and as simple as that for two-body decays.
In the future we shall discuss more delicate issues, such as
CP asymmetries \cite{LGT}, phase shifts from meson-meson scattering
\cite{LZL}, and interference effects among different resonances
\cite{BM}.

We thank S. Brodsky, H.Y. Cheng, M. Diehl and A. Garmash for useful
discussions. This work was supported in part by the National Science
Council of R.O.C. under Grant No. NSC-91-2112-M-001-053, by the National
Center for Theoretical Sciences of R.O.C., and by Theory Group of KEK,
Japan.

\begin{figure}[tbp]
\vspace{1cm} \centerline{ \psfig{figure=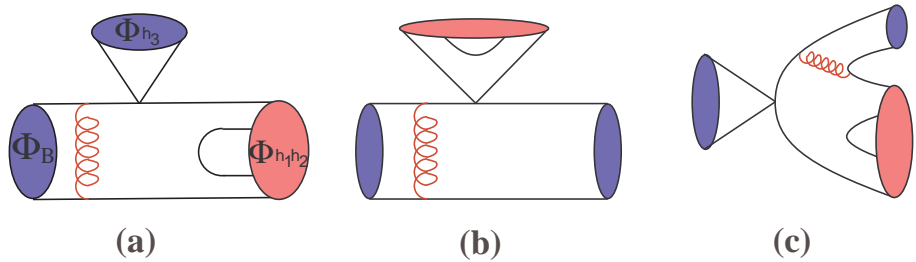,height=1.5in } }
\caption{Graphic definitions for topologies I, II(s), and III.}
\end{figure}

\begin{figure}[tbp]
\vspace{2cm} \centerline{ \psfig{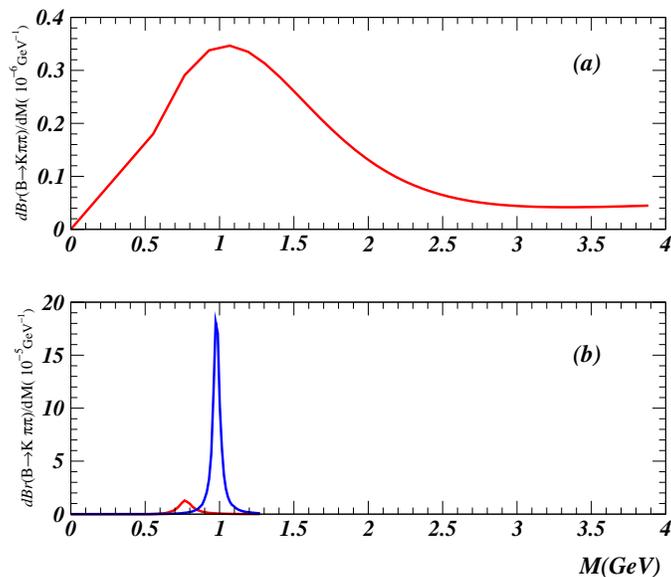} }
\caption{(a) [(b)] Nonresonant (resonant) contribution to the
$B^+\to K^+\pi^+\pi^-$ decay spectrum with respect to the two-pion
invariant mass $M(\pi^+\pi^-)$. The sharp peak corresponds to the
$f_0$ resonance with the width 50 MeV.}
\end{figure}

\end{document}